\documentclass[rnote]{aa} % for the research notes
\usepackage{graphicx}
\usepackage{natbib}

\usepackage{txfonts}
\begin{document}
 \title{X-ray emission from the remarkable A-type star HR 8799}

   \author{J. Robrade
          \and
          J.H.M.M. Schmitt
          }

%   \offprints{J. Robrade}

        \institute{Universit\"at Hamburg, Hamburger Sternwarte, Gojenbergsweg 112, D-21029 Hamburg, Germany\\
       \email{jrobrade@hs.uni-hamburg.de}
             }

   \date{Received...; Accepted...}

% \abstract{}{}{}{}{} 
% 5 {} token are mandatory
 
  \abstract
{}
{A strong decline of magnetic activity towards hotter stars occurs in the regime of mid/late A-type stars due to the vanishing of the outer convection zone.
X-ray emission is an important diagnostic of studying possible activity in intermediate-mass stars.}
{We present a Chandra observation of the exceptional planet bearing A5\,V star HR~8799, more precisely classified as a kA5\,hF0\,mA5 star and search for intrinsic X-ray emission.}
{We clearly detect HR~8799 at soft X-ray energies with the ACIS-S detector in a 10~ks exposure;
minor X-ray brightness variability is present during the observation.
The coronal plasma is described well by a model with a
temperature of around 3\,MK and an X-ray luminosity of about $L_{\rm X} = 1.3 \times 10^{28}$~erg/s in the 0.2\,--\,2.0~keV band,
corresponding to an activity level of log~$L_{\rm X}$/$L_{\rm bol}\approx -6.2$. 
Altogether, these findings point to a rather weakly active and given a RASS detection, long-term stable X-ray emitting star.}
{The X-ray emission from HR~8799 resembles those of a late A/early F-type stars, in agreement with its
classification from hydrogen lines and effective temperature determination and thus resolving the apparent discrepancy with the standard picture of magnetic activity that predicts
mid A-type stars to be virtually X-ray dark.}

   \maketitle
%
%________________________________________________________________

\section{Introduction}

The star \object{HR~8799} (V\,342 Peg, HD~218396) is usually classified as spectral type A5\,V, located at a distance of 39.9~pc and
has a visible magnitude of V\,=\,5.96~mag with B-V\,=\,0.23.
It is an exceptional and so far unique star, it is a $\lambda$~Bootis, $\gamma$\,Doradus and Vega-like at the same time.
$\gamma$\,Doradus stars are a class of pulsating stars
that typically reside at the cool edge of the Cepheid instability strip with spectral types mid-A to mid-F.
HR~8799 was linked to the class of $\lambda$~Bootis stars
by \cite{gray99}; $\lambda$~Bootis stars are chemically peculiar, metal-poor
(in particular the Fe-peak element,s but with the exception of C, N, O) A-type stars that do not exhibit ordered large-scale magnetic fields as observed in Ap stars \citep{boh90}.
\cite{gray99} derived a metallicity of [M/H]\,=\,-0.47 for HR~8799 and determined its stellar parameters to 
$M\,=\,1.47\pm 0.30$\,M$_{\odot}$, $R\,=\,1.34\pm 0.05$\,R$_{\odot}$, $L\,=4.92\pm 0.41$\,L$_{\odot}$ ($1.9 \times 10^{34}$~erg/s) and
$T_{\rm eff}=7430 \pm 75$~K.
They assigned it the spectral type kA5\,hF0\,mA5\,v\,$\lambda$\,Boo to indicate that HR~8799 is a mild $\lambda$\,Bootis star that exhibits the A5 standard in the CaII\,K line
as well as in other metallic lines, however the hydrogen profiles (and thus effective temperature) are in better agreement with the F0 standard. 
The $\lambda$\,Bootis phenomenon is thought to be caused by the accretion of metal depleted material, i.e. external processes, in line with the finding that
HR~8799 is also a 'Vega-like' star that shows a far-IR excess flux at 60~$\mu$m from circumstellar dust in IRAS data \citep{sad86}.
Additionally, there is strong evidence for a dusty debris disk from IR observations \citep{zuck04}.

Recently, HR~8799 gained even more interest when multiple orbiting planets were detected by direct imaging \citep{mar08}.
The three planets are located at distances of several tens (24, 38, 68) AU and have masses around 10~$M_{\rm J}$ each.
The authors derived an age of about 60\,Myr (30\,--\,150 Myr) from various lines of evidence, 
thus HR~8799 is likely relatively young.
While stellar radiation, especially energetic ones like UV and X-rays, 
influences the evolution of circumstellar material and thus planet formation, 
the planets themselves are likely not important in the generation of X-ray emission in such stellar systems.
Although its $Vsini=38 \pm2$~km/s \citep{kaye98}
is only moderate for an A-type star and the exact orientation of the rotation axis is unknown,
astrometric analysis of the detected planetary system suggests a rather low inclination 
and consequently fast rotating star, e.g. $V_{\rm rot} \gtrsim$~100 (200)~km/s for $i \lesssim 20~(10)^{\circ}$.

Hints for X-ray emission from HR~8799 came from RASS (ROSAT All-Sky-Survey) data via a cross-correlation search between X-ray positions 
and bright A-type stars \citep{hue98}.
They found that the position of the soft X-ray source 1RXS~J230729.0+210802 matches quite well with the one of HR~8799;
however, the positional uncertainty and the low number of detected counts (10 photons) prevent a detailed investigation.
On the other hand, \cite{kaye98} attributed the chromospheric \ion{Ca}{ii}~K flux of HR~8799 to be of basal origin.
From the X-ray point of view HR~8799 is remarkable, since an A5 star should be virtually X-ray dark in the standard paradigm
that is valid for main sequence stars.
In this scheme X-ray emission is expected from magnetic activity in the regime of 'cool stars' (spectral type around A7 and later) 
and from wind-shocks in the regime of 'hot stars' (B2 and earlier); intermediate spectral types are virtually X-ray dark.
Exceptions are found among young Herbig AeBe stars
and peculiar Ap/Bp stars, where fossil magnetic fields are thought to play a mayor role in the generation of X-rays.
This picture is well established by X-ray observations 
with {\it Einstein} and ROSAT,
showing that stellar activity develops in late A-type stars and increases strongly towards
cooler stars, with activity levels being in the range of $\log L_{\rm X}$/$L_{\rm bol}= -3~...~-7$ \citep[e.g.][]{schmitt85, schmitt97}.
A recent {\it XMM-Newton} observation of the fast rotating A7 star Altair confirmed the presence of weak magnetic activity and
coronal X-ray emission \citep{rob09a}.
The vanishing of magnetic activity in mid A-type stars is expected theoretically and
confirmed with other activity indicators; e.g. FUSE observations of main sequence stars have shown the disappearance of chromospheric emission
lines at effective temperatures above $T_{\rm eff} \approx 8200$\,K \citep{sim02}.

In this paper we present results from a {\it Chandra} ACIS-S observation of HR~8799. Our paper is structured as follows:
in Sect.\,\ref{ana} we describe the observation and data analysis, in Sect.\,\ref{res} we present our results
and summarize our findings in Sect.\,\ref{sum}.

\section{Observations and data analysis}
\label{ana}

The target HR~8799 was observed by {\it Chandra} with the back-illuminated S3 chip of the 
\hbox{ACIS-S} detector in August~2009 for about 10\,ks (Obs.-ID 10975).
We use the standard software package CIAO~4.1 to analyze the data, including the
tool 'wavedetect' to derive the source position.
Source photons were found in the 0.15\,--\,2.0~keV energy range, where we detect 137 photons in a 2\,\arcsec\,circular
region around the position of HR~8799 with an expected background of 0.2 photons as deduced from nearby source-free regions.
These photons are denoted as source photons in the following and
enable a more detailed study of the X-ray properties of HR~8799, compared to the ROSAT data.

Spectral analysis is done with XSPEC~V12.3 \citep{xspec} and we require a minimum of 15 counts per spectral bin in our modelling.
To model the spectra we
use photo-absorbed, multi-temperature APEC models; all abundances are relative to solar values as given by \cite{grsa}.
We find that no absorption component is required to describe the obtained X-ray spectrum, enforcing interstellar absorption
at a level of $N_{H} = 10^{19} - 10^{20}$~cm$^{-2}$ does not significantly change the results.

\begin{figure}[ht]
\hspace*{1mm}
\includegraphics[width=42mm]{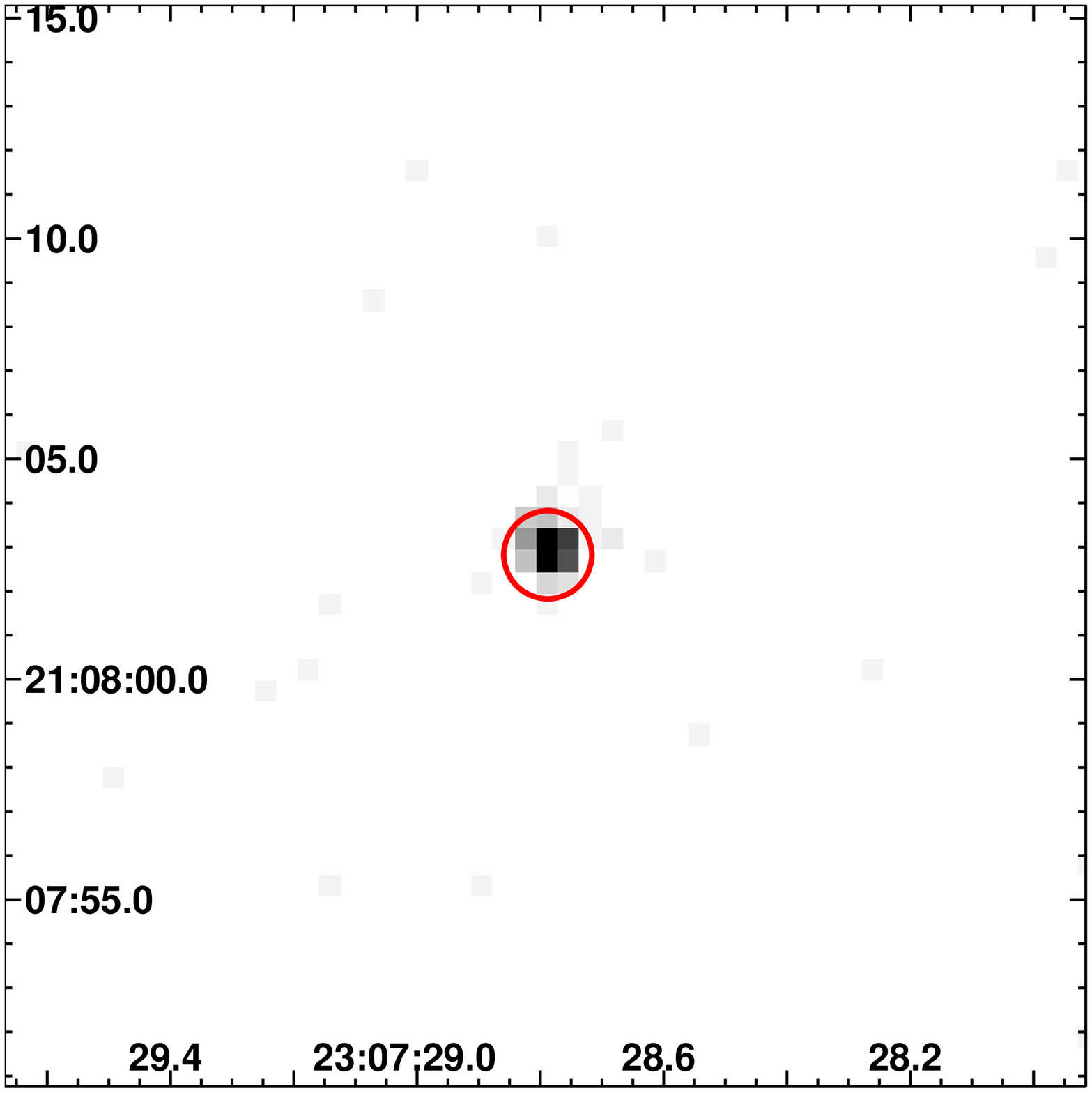}
\hspace*{2mm}
\includegraphics[width=42mm]{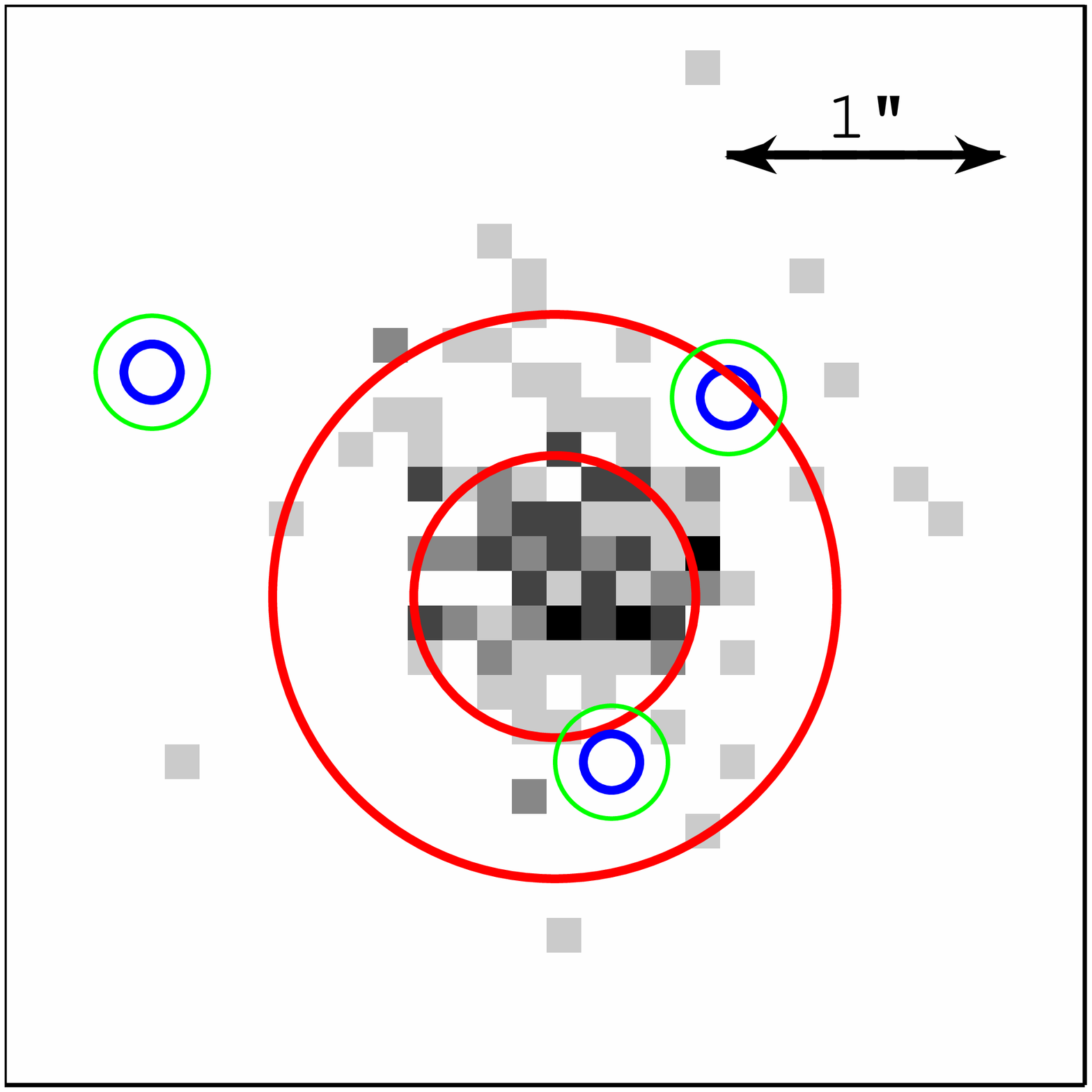}
\caption{\label{im}{\it Left}: ACIS-S image of HR~8799 and optical position.
{\it Right}: Sub-pixel image with the three planets marked (see text for details).}
\end{figure}

\section{Results}
\label{res}

\subsection{Images and light curves}

In the left panel of Fig.\,\ref{im} we show the obtained X-ray image of the source and the optical position of HR~8799,
denoted by a circle with a 1\arcsec\,radius, corresponding to $\approx 95$\% encircled X-ray energy.
The right panel shows a zoom-in of the source, the outer circle is the same as in the left panel, whereas the
inner half-sized circle corresponds to 65\% encircled energy.
Additionally, the planetary positions are marked (small circles with 0.1\arcsec~and 0.2\,\arcsec~radius).
Only one X-ray source is present in the vicinity; 
it appears point-source like and its position agrees with an absolute offset of only 0.2\arcsec\,
very well with the optical position of the A-type star HR~8799. 
No significant excess
emission is seen at any of the positions of the detected planets (separations to HR~8799 between 0.6\arcsec~and 1.7\arcsec)
nor at the positions of the substellar companions that were detected as point-source candidates
in a coronographic survey performed with NICMOS/HST at distances of 13.7~\arcsec and 15.7~\arcsec \citep{low05}.
Possible X-ray emission from any of these objects is at least two orders of magnitude fainter than those from
HR~8799. HR~8799 is thought to be a single star and given the accurate source position obtained from the ACIS data
we attribute the detected X-ray emission exclusively to the A5 star HR~8799.

\cite{boh90} put upper limits of a few hundred Gauss on the longitudinal magnetic field component in a sample of $\lambda$~Bootis stars, not including HR~8799
and ruling out the existence of strong, ordered fields. However, small-scale magnetic fields as expected for a corona that originates 
from magnetic activity may easily be present.
The X-ray emission from HR~8799 implies the presence of a magnetic field on HR~8799, thus $\lambda$~Bootis stars are not exclusively non-magnetic.

\begin{figure}[t]
\hspace*{1.5mm}
\includegraphics[width=91mm]{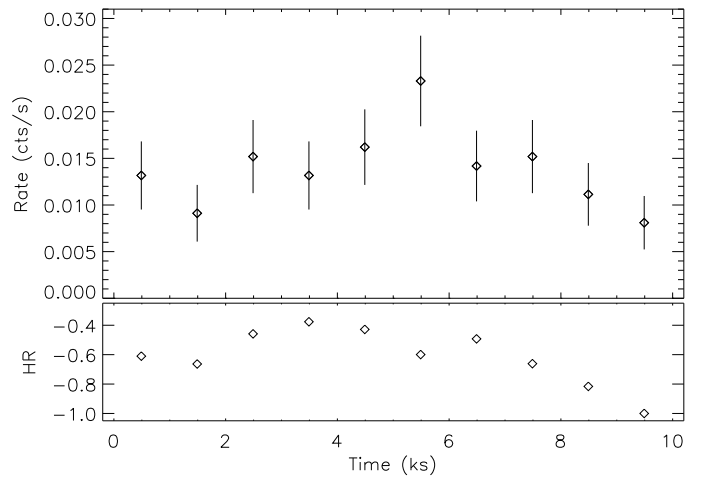}
\caption{\label{lc}Light curve and hardness ratio of HR~8799 (1~ks binning).}
\end{figure}

To investigate the X-ray variability of HR~8799,
we create a light curve with 1\,ks binning from the source photons (Fig.\,\ref{lc}, upper panel).
The light curve is moderately variable; the count rate changes by up to a factor of about two, 
however the exact level of the variability is not well constrained.
We also studied light curves with shorter time bins of e.g. 100\,s, but find no indications for strong short-term variability that would be
present for burst-like flares.

We study the origin of the observed variability and
search for spectral variations related to the changes in X-ray brightness by determining the respective hardness ratio
$HR=(H-S)/(H+S)$ for each time bin, with $S$\,=\,0.15\,--\,0.7~keV and $H$\,=\,0.7\,--\,2.0~keV being the respective 
photon energy bands (see lower panel, Fig.\,\ref{lc}).
Spectral lines are unresolved in the ACIS spectrum, however the emission 
below 0.7~keV is dominated by cooler plasma with line peak formation temperatures in the range of 1\,--\,4~MK, while above 0.7\,keV lines
from hotter plasma ($\gtrsim 5$~MK) dominate. If enhanced magnetic activity is the origin for the X-ray brightening, 
one expects a correlation of the X-ray brightness with the hardness of the emission.
The error on the individual hardness ratio bins is about $\pm 0.25$, but indeed
harder emission is observed during X-ray brighter phases. A Spearman's rank correlation gives a chance probability
of 0.07 that the correlation is due to a statistical fluctuation for the used ten time bins.
Thus variable magnetic activity is the likely cause for the observed moderate X-ray brightness variability, whereas more extreme
flaring events as observed on active stars are not present on HR~8799 during our observation.

\subsection{Spectral analysis}

\begin{figure}[t]
\hspace*{2mm}\includegraphics[width=55mm,angle=-90]{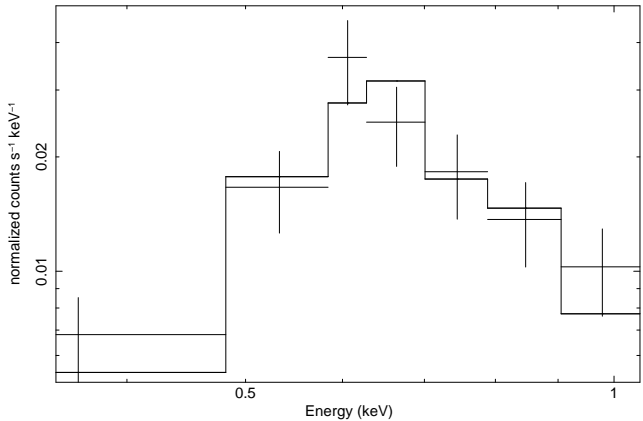}
\caption{\label{spec}ACIS-S spectrum of HR~8799 with applied spectral model.}
\end{figure}

We determine the global spectral properties of HR~8799 by fitting the ACIS spectrum
with multi-temperature spectral models.
The metallicity of the coronal plasma can only be poorly constrained with the existing data and was fixed. Using the
photospheric value of HR~8799, determined to 0.33 solar abundances \citep{gray99},
already a one temperature model is sufficient to describe the data.
The corona therefore also seems to exhibit the photospheric sub-solar composition, but 
Adopting solar abundances results in correspondingly lower emission measure, albeit
the quality of the fit is slightly poorer for a one temperature model.
The corona therfore is better described by the photospheric sub-solar composition, but 
the indications are sparse; we note
that a similarly good fit can be obtained with solar metallicity and two, however poorly constrained, temperature components.
We show the X-ray spectrum and the best fit one temperature model in Fig.\,\ref{spec},
the derived spectral properties are summarized in Table\,\ref{fit}.

From the best fit model we obtain an
X-ray luminosity of $L_{\rm X}= 1.3 \times 10^{28}$~erg/s in the 0.2\,--\,2.0~keV band, corresponding to
a surface flux of $\log F_{\rm X}= 5.1$~erg\,cm$^{-2}$\,s$^{-1}$ and an 
activity level of log~$L_{\rm X}$/$L_{\rm bol} = -6.16$, indicating that HR~8799 is despite its youth a rather weakly active star.
Since HR~8799 is about an order of magnitude X-ray brighter than the average Sun, the X-ray flux at the position of the closest of the detected planets
corresponds to the solar X-ray flux that is received by Saturn.
To compare our result with the RASS measurement, we also determine the X-ray luminosity for the ROSAT 0.1\,--\,2.4\,keV band
from the spectral model, leading to a value of $L_{\rm X}= 1.6 \times 10^{28}$~erg/s.
The ROSAT X-ray luminosity of $L_{\rm X} = 1.5 \times 10^{28}$~erg/s, obtained from a hardness dependent energy conversion factor, is basically the same as the
one obtained with {\it Chandra} more than 25~years later, pointing to the presence of long-term stable X-ray emission.

\begin{table}[t]
\begin{center}
\caption{\label{fit} Spectral fit results derived from ACIS-S data.}
\begin{tabular}{lrrl}
\hline\hline
Par.  &  \multicolumn{2}{c}{1-T models}  &unit\\\hline
T1& 0.26$\pm$0.03 & 0.26$\pm$0.03 & keV\\
Abund.&0.33 & 1.0 &solar\\
EM1&1.9$\pm0.3 \times 10^{51}$ &6.9$\pm1.1 \times 10^{50}$&cm$^{-3}$ \\\hline
$\chi^2_{red}${\tiny(d.o.f.)}& 0.83 (5) & 1.14 (5) &\\\hline
\end{tabular}
\end{center}
\end{table}

\subsection{HR 8799 in the context of A-type stars}

The activity level as well as the X-ray luminosity of HR~8799 are about an order of magnitude larger than 
those of magnetically active mid to late A-type stars like Alderamin ($\alpha$~Cep) or Altair ($\alpha$~Aql), both of
spectral type A7.
These stars have X-ray luminosities of one to a few times $10^{27}$~erg/s and activity levels of
log~$L_{\rm X}$/$L_{\rm bol}\lesssim -7$.
Compared to Altair, the hottest magnetically active star studied in detail at X-ray energies \citep{rob09a},
the X-ray surface flux of HR~8799 is by about a factor 20 higher; further the
average coronal temperature of HR~8799 is with 3.0~MK slightly higher than those of Altair that is around 2.5~MK,
but within the errors this difference is not very significant.
This matches the finding that Altair's activity level is with log~$L_{\rm X}$/$L_{\rm bol} = -7.4$ very low,
even when compared to the weakly active stars HR~8799.
However, the ultra-fast rotator Altair is already around its maximum possible activity level 
and a significant spin-up would disrupt the star.
Consequently, the dynamo mechanism in HR~8799 needs to be more efficient than that in Altair.
The efficiency of a solar-type dynamo is proportional to the 
inverse square of the Rossby-Number (Ro\,=\,$P/\tau_{c}$), with $P$ denoting the rotational period and $\tau_{c}$ the
convective turnover time. A significantly shorter rotation period for HR~8799 is ruled out when considering stellar dimensions,
$Vsini$ and geometry; thus
a deeper convection zone would provide a natural explanation for the required efficiency.

This is in line with the finding that for the spectral classification as an A5 star
the determined $T_{\rm eff}\approx$\,7400~K from \cite{gray99} is rather low.
Thus one might assume that its internal structure is better described by a F0 star. In this case the outer
convection zone would easily provide sufficient dynamo action to generate the observed X-ray emission from magnetic activity.
The still rather low activity level of log~$L_{\rm X}/L_{\rm bol} \approx -6.1$
(active late-type stars have log~$L_{\rm X}/L_{\rm bol} \approx -3$) that is required 
to generate the observed X-ray emission from HR~8799 is easily achieved in this case.
Further, as it is also the case for Altair, the surface of HR~8799 is not necessarily homogeneous 
and specific surface areas like an equatorial bulge may contribute predominantly to the X-ray emission 

The activity level as well as the the coronal properties support the hypothesis that the often used 'metallic'
classification as A5 star does not reflect the properties that underly magnetic activity phenomena.
From the X-ray point of view, HR~8799 resembles much more
an late A or early F-type star. This is in line with its classification based on hydrogen lines and its effective temperature derived from spectral modelling.
Consequently, the X-ray emission from HR~8799 appears rather natural. Sufficient sensitive searches with
other indicators should also reveal signs of magnetic activity phenomena on this remarkable star.

\section{Summary and conclusions}
\label{sum}

   \begin{enumerate}
\item We have detected soft X-ray emission from the extraordinary A5/F0 V star HR~8799.
The derived X-ray luminosity of about $L_{\rm X} = 1.3 \times 10^{28}$~erg/s in the 0.2\,--\,2.0~keV band corresponds
to an activity level of log~$L_{\rm X}$/$L_{\rm bol}\approx -6.2$, pointing to a weakly active star.

\item The ACIS data confirms the ROSAT detection, significantly improves the positional accuracy
and additionally enables a deeper study of the X-ray properties of HR~8799. We find that
its X-ray emitting coronal plasma has an average temperature of about 3.0~MK, typical for a star with a rather low activity level.
Minor X-ray brightness variations
are present in our observation; the accompanying spectral changes point to variable magnetic activity.

\item Overall, the X-ray properties of HR~8799 resemble those of mildly active late A/early F stars, rather than those of
mid/late A-type stars. Keeping in mind that the attributed spectral type for HR~8799 depends on
the classification criterion, the one based on hydrogen lines (F0) most suitable reflects its X-ray properties.

   \end{enumerate}

\begin{acknowledgements}
This work is based on observations obtained with {\it Chandra}.
J.R. acknowledges support from DLR under 50QR0803.

\end{acknowledgements}

\bibliographystyle{aa}
\bibliography{14027}

\begin{thebibliography}{14}
\expandafter\ifx\csname natexlab\endcsname\relax\def\natexlab#1{#1}\fi

\bibitem[{{Arnaud}(1996)}]{xspec}
{Arnaud}, K.~A. 1996, in ASP Conf. Ser. 101: Astronomical Data Analysis
  Software and Systems V, ed. G.~H. {Jacoby} \& J.~{Barnes}, 17

\bibitem[{{Bohlender} \& {Landstreet}(1990)}]{boh90}
{Bohlender}, D.~A. \& {Landstreet}, J.~D. 1990, \mnras, 247, 606

\bibitem[{{Gray} \& {Kaye}(1999)}]{gray99}
{Gray}, R.~O. \& {Kaye}, A.~B. 1999, \aj, 118, 2993

\bibitem[{{Grevesse} \& {Sauval}(1998)}]{grsa}
{Grevesse}, N. \& {Sauval}, A.~J. 1998, Space Science Reviews, 85, 161

\bibitem[{{H{\"u}nsch} {et~al.}(1998){H{\"u}nsch}, {Schmitt}, \&
  {Voges}}]{hue98}
{H{\"u}nsch}, M., {Schmitt}, J.~H.~M.~M., \& {Voges}, W. 1998, \aaps, 132, 155

\bibitem[{{Kaye} \& {Strassmeier}(1998)}]{kaye98}
{Kaye}, A.~B. \& {Strassmeier}, K.~G. 1998, \mnras, 294, L35

\bibitem[{{Lowrance} {et~al.}(2005){Lowrance}, {Becklin}, {Schneider},
  {Kirkpatrick}, {Weinberger}, {Zuckerman}, {Dumas}, {Beuzit}, {Plait},
  {Malumuth}, {Heap}, {Terrile}, \& {Hines}}]{low05}
{Lowrance}, P.~J., {Becklin}, E.~E., {Schneider}, G., {et~al.} 2005, \aj, 130,
  1845

\bibitem[{{Marois} {et~al.}(2008){Marois}, {Macintosh}, {Barman}, {Zuckerman},
  {Song}, {Patience}, {Lafreni{\`e}re}, \& {Doyon}}]{mar08}
{Marois}, C., {Macintosh}, B., {Barman}, T., {et~al.} 2008, Science, 322, 1348

\bibitem[{{Robrade} \& {Schmitt}(2009)}]{rob09a}
{Robrade}, J. \& {Schmitt}, J.~H.~M.~M. 2009, \aap, 497, 511

\bibitem[{{Sadakane} \& {Nishida}(1986)}]{sad86}
{Sadakane}, K. \& {Nishida}, M. 1986, \pasp, 98, 685

\bibitem[{{Schmitt}(1997)}]{schmitt97}
{Schmitt}, J.~H.~M.~M. 1997, \aap, 318, 215

\bibitem[{{Schmitt} {et~al.}(1985){Schmitt}, {Golub}, {Harnden}, {Maxson},
  {Rosner}, \& {Vaiana}}]{schmitt85}
{Schmitt}, J.~H.~M.~M., {Golub}, L., {Harnden}, Jr., F.~R., {et~al.} 1985,
  \apj, 290, 307

\bibitem[{{Simon} {et~al.}(2002){Simon}, {Ayres}, {Redfield}, \&
  {Linsky}}]{sim02}
{Simon}, T., {Ayres}, T.~R., {Redfield}, S., \& {Linsky}, J.~L. 2002, \apj,
  579, 800

\bibitem[{{Zuckerman} \& {Song}(2004)}]{zuck04}
{Zuckerman}, B. \& {Song}, I. 2004, \apj, 603, 738

\end{thebibliography}

\end{document}